# High-sensing properties of magnetic plasmon resonances in double- and triple-rod structures

J. X. Cao[1], H. Liu[1,*], T. Li[1], S. M. Wang[1], Z. G. Dong[2], L. Li[1], C. Zhu[1], Y. Wang[1], and S. N. Zhu[1,*]

[1]Department of Physics, National Laboratory of Solid State Microstructures, Nanjing University, Nanjing 210093, People's Republic of China
[2]Physics Department, Southeast University, Nanjing 211189, People's Republic of China
*zhusn@nju.edu.cn; *liuhui@nju.edu.cn

**Abstract**: We numerically investigated the magnetic plasmon resonances in double-rod and triple-rod structures (DRSs and TRSs, respectively) for sensing applications. According to the equivalent circuit model, one magnetic plasmon mode was induced in the DRS. Due to the hybridization effect, two magnetic plasmon modes were obtained in the TRS. Compared with the electric plasmon resonance in a single-rod structure (SRS), the electromagnetic fields near the DRS and TRS were much more localized in the dielectric surrounding the structures at the resonance wavelengths. This caused the magnetic plasmon resonance wavelengths to become very sensitive to refractive index changes in the environment medium. As a result, a large figure of merit that is much larger than the electric plasmon modes of SRS could be obtained in the magnetic plasmon modes of DRS and TRS. These magnetic plasmon mode properties enable the use of DRSs and TRSs as sensing elements with remarkable performance.

## 1. Introduction

Optical sensors based on the excitation of surface plasmon resonance in nano-metallic structures have been greatly developed in recent years due to their great local field enhancement effect [1-8]. Various shapes of metal nanoparticles have been proposed, such as nanorods [9], nanoshells [10], nanocages [11], and nanostars [12]. Recently, Piliarik et al. predicted the ultimate performance of all major configurations of surface plasmon resonance sensors [13]. Double-rod structures (DRSs) were proposed by Shalaev et al. to realize magnetic plasmon resonance existent in split-ring resonators [14] and negative refraction in infrared frequency ranges [15]. In our previous work [16], a triple-rod structure (TRS) was reported. This structure could be seen as two coupled DRSs. Two orthogonal eigen magnetic plasmon modes could be excited in the TRS due to its strong hybridization effect. Omni-directional negative refraction [16] and great polarization changes [17] could also be realized in TRSs.

In this paper, the magnetic plasmon modes in the DRS and TRS were employed to realize high-quality sensing. We investigated the sensing properties of the magnetic plasmon modes in the DRS/TRS and compared them with those of the electric plasmon modes in the SRS. Based on the simulations, the electromagnetic fields were greatly confined to the dielectric surrounding the designed structures at the magnetic plasmon resonance wavelengths, indicating that the magnetic plasmon modes were highly sensitive to refractive index changes in the dielectric. Together with the sharp dip and narrow full-width at half-maximum (*FWHM*) of the transmission curves, the magnetic plasmon modes of the DRS and TRS showed sensing performance far superior to that of the electric plasmon modes in the SRS. In particular, the short wavelength hybrid mode in TRS showed a figure of merit (*FOM*) of 96.3, which was larger than that of other structures investigated in this paper.

## 2. Design of numerical models

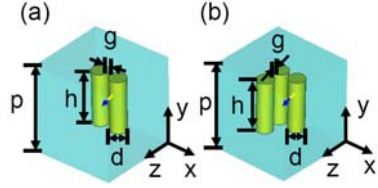

Fig. 1. Schematics of unit cell for (a) DRS and (b) TRS with related geometry parameters, h = 300nm, g = 20nm, d = 100nm, and the period p = 500nm. The blue arrows in the figures represented the probes detecting magnetic field component |H$_z$|.

Figure 1 represents the schematic of the unit cell for the DRS and TRS. Both two-cube unit cells had the same length of side p = 500 nm. The length and cross-section diameter of the gold rods were h = 300 nm and d = 100 nm, respectively. The gap between the rods was g = 20 nm. The boundary of the cells for the DRS and TRS in the x- and z-directions was defined as a periodic condition; that in the y-direction was set as an open condition. We first set the medium surrounding the gold rods as air ($n$ = 1) and studied the magnetic responses and transverse field distributions. Afterwards, we changed the refractive index of the surrounding medium to investigate the enhanced sensing characteristics.

To study the magnetic plasmon characters of the proposed structures, the commercial software package CST Microwave Studio (Computer Simulation Technology GmbH, Darmstadt, Germany) was employed in the numerical analysis. The permittivity of gold was satisfied by the Drude model $\varepsilon(\omega) = 1 - \omega_p^2/(\omega^2 + i\omega_\tau\omega)$, with plasma frequency $\omega_p$ = 9.02 eV and damping constant $\omega_\tau$ = 0.027 eV [18]. Considering the grain boundary effect, surface scattering, and inhomogeneous broadening in the thin film, the damp constant used in the calculations was approximately three times as large as that of bulk gold [7].

### 3. Results and discussions

In the simulations, the propagation direction of the linearly polarized incident wave was along the y-axis (Fig. 3). The polarization direction was defined by the angle $\theta$ between the electric field and the x-axis. For the DRS, the magnetic resonance mode was excited by the incident wave with $\theta$ = 0° (Figs. 3(a)-3(b)). According to our previous work [17], two hybrid magnetic plasmon modes could be established for the TRS. When $\theta$ = 45° (Figs. 3(c)-3(d)), only the long wavelength mode (LW mode) was excited at $\lambda_+$; when $\theta$ = -45° (Figs. 3(e)-3(f)), only the short wavelength mode (SW mode) was excited at $\lambda_-$. In this paper, the magnetic response of the TRS was excited by the incident wave with $\theta$ = 45° and -45°. Figure 2 shows the magnetic amplitude |H$_z$| recorded by the probes located inside the gaps between the nanorods (shown in Fig. 1). The resonance wavelength for the DRS was observed at $\lambda_0$ = 1.167 μm; while for the TRS, the resonance wavelengths were at $\lambda_+$ = 1.208 μm and $\lambda_-$ = 1.121 μm. To study the resonance properties of the three magnetic plasmon modes, the $Q$ factor of the DRS and TRS was calculated based on the definition $Q = (2\pi\lambda)/\Delta\lambda$, where $\lambda$ and $\Delta\lambda$ were the resonance wavelengths and *FWHM* of the resonance peak, respectively. From our simulation results, the calculated $Q$ factor of the magnetic plasmon mode in the DRS was $Q_0$ = 806.87. For the LW and SW modes in the TRS, the $Q$ factors were $Q_+$ = 764.99 and $Q_-$ = 934.21. Obviously, the $Q$ factor of the SW mode in TRS was the best.

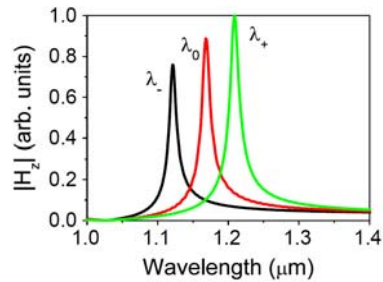

Fig. 2. Magnetic field amplitude |H$_z$| detected by the probes shown in Fig. 1.

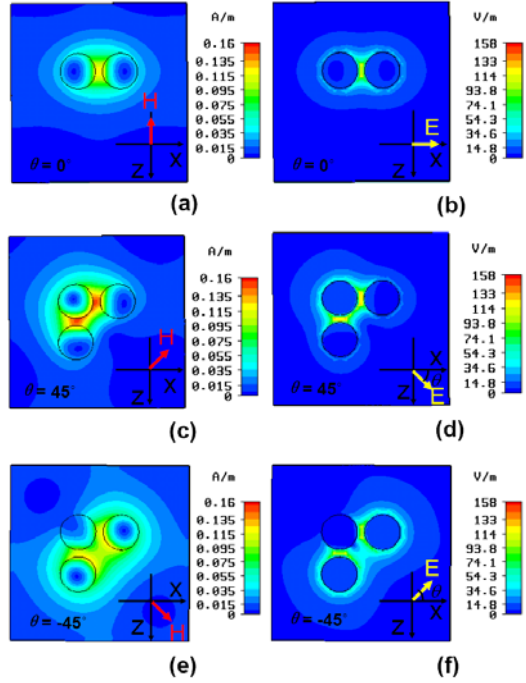

Fig. 3. Magnetic field distributions and electric field patterns at magnetic plasmon resonance wavelengths (a)-(b) $\lambda_0$ for DRS, (c)-(d) $\lambda_+$, and (e)-(f) $\lambda_-$ for TRS, respectively. The red arrows and the yellow ones in the figures showed the directions of the incident magnetic and electric fields.

Figure 3 shows the magnetic field distributions in the middle section plane (y=150 nm) and the electric field patterns in the end plane (y=300 nm) of the nanorods at $\lambda_0$ (Figs. 3(a)-3(b)) for the DRS, and $\lambda_+$ (Fig. 3(c)-(d)) and $\lambda_-$ (Fig. 3(e)-(f)) for the TRS. For the DRS, the electric field was localized around the rods. In particular, the magnetic field was nearly confined inside the gap between the two rods, which showed much more confinement than the electric field. For the TRS, the two induced currents oscillated along the same direction in the middle rod for the LW mode but in opposite directions for the SW mode [17]. As a result, the electric field circled the middle-rod at $\lambda_+$ (Fig. 3(c)-3(d)) and circled the two side-rods at $\lambda_-$ (Fig. 3(e)-3(f)). In order to provide good comparison, the plasmon resonance of SRS was also investigated. Two different plasmon modes could be excited in SRS under different incident polarization directions (Fig. 4). The longitudinal mode (LLSP mode) was excited by the electric field along the rod (Fig. 4(a)), and the transverse mode (TLSP mode) was excited by the electric field perpendicular to the rod (Fig. 4(b)). The comparison between Figs. 3 and 4 shows that the enhancement of the local electric field in the TRS was better than in both the SRS and DRS.

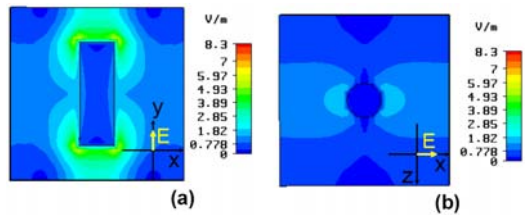

Fig. 4. Electric field patterns at (a) LLSP and (b) TLSP wavelengths in the SRS. The yellow arrows in the figures represented the direction of the incident electric field.

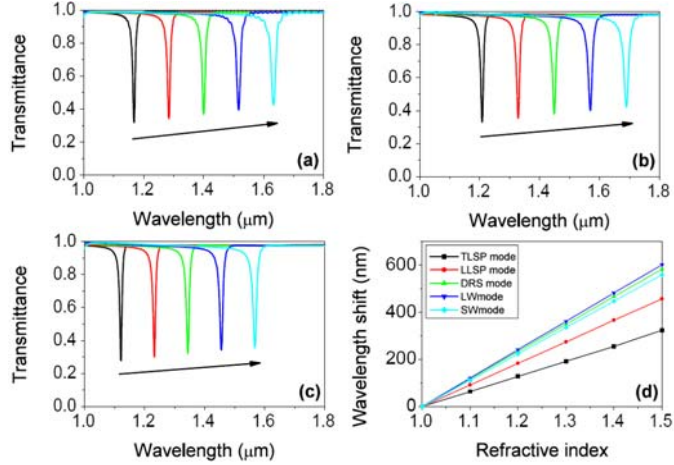

Fig. 5. Transmissions of metamaterials based on (a) DRSs and (b)-(c) TRSs with the incident cases shown in Fig. 3. (d) Dependence of resonant wavelength shifts to the refractive index of the dielectric surrounding the DRSs and TRSs, together with that in the SRSs.

Due to the intensive localization of electromagnetic energy inside the proposed structures, the resonance wavelengths of the magnetic plasmon modes were very sensitive to the refractive index change in the surrounding materials. Thus, the DRS and TRS could be used as highly sensitive sensors. Figure 5 shows the dependence of the transmission and resonance wavelength shifts on the refractive index of the surrounding materials. From Figs. 5(a)-5(c), when the refractive index of the surrounding dielectric was varied from 1.0 to 1.4, the resonance wavelength of DRS shifted from 1.167 to 1.634 μm. For the TRS, the resonance wavelength shifted from 1.208 to 1.689 μm for the LW mode, and from 1.121 to 1.567 μm for the SW mode. Figure 5(d) shows that the resonance wavelength shifts of the magnetic plasmon modes were perfectly linear over the whole refractive index range. Sensitivity, denoted as the slope of the resonance wavelength over the refractive index change, was 1160.3 nm/refractive index unit (RIU) for the magnetic plasmon mode of the DRS. For the LW mode of the TRS, sensitivity was 1208.8 nm/RIU and larger than DRS. For the SW mode, sensitivity was 1114.6 nm/RIU and smaller than DRS. To demonstrate the advantage of the magnetic plasmon modes for sensing, we also calculated the sensitivity of the electric plasmon resonances of SRS. Sensitivity of the LLSP mode was 912.7 nm/RIU, and 633.8 nm/RIU for the TLSP mode. Both of the sensitivities of the electric plasmon modes were much smaller than those of the magnetic plasmon modes of the DRS and TRS.

To directly compare the overall sensing performance of the electric plasmon modes in SRSs and magnetic plasmon modes in DRSs and TRSs, a useful concept of *FOM* was used, defined as follows [1, 6]:

$$FOM = \frac{m(nm\ RIU^{-1})}{FWHM(nm)}(1-T_{\min}),\qquad(1)$$

where $m$ is the slope of the resonance wavelength over the refractive index change. The calculated $m$, *FWHM*, $T_{\min}$, and *FOM* values are given in Table I. It could be seen that the magnetic plasmon modes showed better sensing performance than the electric plasmon modes due to the high sensitivity and strong capability of electromagnetic field localization. For instance, the *FOM* for the SW mode was 96.3, approximately 21.7% greater than magnetic plasmon mode in the DRSs. It was also more than 35 times that of the LLSP mode in the SRSs. According to our previous semi-analytic theory [17-19], the imaginary part of the permeability element for the SW mode was smaller than that of the LW mode of TRSs and magnetic plasmon mode of DRSs. Therefore, the SW mode had a higher quality factor. As a result, the best *FOM* could still be obtained in the SW mode.

**Table I.** Calculations of sensing performance by figure of merit (*FOM*) and the related parameters. Obviously, the SW mode in the TRS had the greatest *FOM* among that of the plasmon modes in the SRSs, DRSs and TRSs.

| Eigenmode | $m$ (nm/RIU) | *FWHM* (nm) | $T_{min}$ | *FOM* |
|---|---|---|---|---|
| LLSP mode of SRSs | 912.7 | 357.24 | $7.81 \times 10^{-5}$ | 2.55 |
| TLSP mode of SRSs | 633.8 | 8.77 | 0.285 | 51.7 |
| MPR of DRSs | 1160.3 | 10.00 | 0.318 | 79.1 |
| LW mode of TRSs | 1208.8 | 10.70 | 0.330 | 75.7 |
| SW mode of TRSs | 1114.6 | 8.37 | 0.277 | 96.3 |

## 4. Conclusion

In conclusion, we numerically investigated the sensing properties of the magnetic plasmon modes of DRSs and TRSs. The electromagnetic field was strongly confined within the proposed structures at the magnetic plasmon modes, leading to greater field-material interactions and increased sensitivity to the refractive index changes of the surrounding medium. In addition to high sensitivity, the transmission for the excitation of the hybrid SW mode of the TRS had a good *Q* factor and possessed a good *FOM* of 96.3. Compared with the electric plasmon modes in the SRSs, the magnetic plasmon modes had higher sensitivity and could offer better performance as sensors in the infrared range.

**Acknowledgements**

This work is supported by the National Natural Science Foundation of China (No.10704036, No.10874081, No.60907009, No.10904012, No.10974090 and No. 60990320), and by the National Key Projects for Basic Researches of China (No. 2006CB921804, No. 2009CB930501 and No. 2010CB630703).

## References

1. L. J. Sherry, S. H. Chang, G. C. Schatz, R. P. Van Duyne, B. J. Wiley, and Y. N. Xia, "Localized surface plasmon resonance spectroscopy of single silver nanocubes," Nano Lett. **5**, 2034-2038 (2005).
2. H. W. Liao, C. L. Nehl, and J. H. Hafner, "Biomedical applications of plasmon resonant metal nanoparticles," Nanomedicine **1**, 201-208 (2006).
3. Hui Wang, Daniel W. Brandl, Fei Le, Peter Nordlander, and Naomi J. Halas, "Nanorice: A hybrid plasmonic nanostructure," Nano Lett. **6**, 827-832 (2006).
4. Jeffrey N. Anker, W. Paige Hall, Olga Lyandres, Nilam C. Shah, Jing Zhao, and Richard P. Van Duyne, "Biosensing with plasmonic nanosensors," Nature materials **7**, 442-453 (2008).
5. Jiří Homola, "Surface plasmon resonance sensors for detection of chemical and biological species," Chem. Rev. **108**, 462-493 (2008).
6. Kyung Min Byun, Sung June Kim, and Donghyun Kim, "Grating-coupled transmission-type surface Plasmon resonance sensors based on dielectric and metallic gratings," Appl. Opt. **46**, 5703-5708 (2007).
7. Na Liu, Lutz Langguth, Thomas Weiss, Jürgen Kästel, Michael Fleischhauer, Tilman Pfau, and Harald Giessen, "Plasmonic analogue of electromagnetically induced transparency at the Drude damping limit," Nature Materials **8**, 758-762 (2009).
8. Alp Artar, Ahmet Ali Yanik, and Hatice Altug, "Fabry-Pérot nanocavities in multilayered plasmonic crystals for enhanced biosensing," Appl. Phys. Lett. **95**, 051105 (2009).
9. H. Wang, T. B. Huff, D. A. Zweifel, W. He, P. S. Low, A. Wei, and J. Cheng, "In vitro and in vivo two-photon luminescence imaging of single gold nanorods," Proc. Natl Acad. Sci. USA **102**, 15752-15756 (2005)
10. S. J. Oldenburg, R. D. Averitt, S. L. Westcott, and N. J. Halas, "Nanoengineering of optical resonances," Chem. Phys. Lett. **288**, 243-247 (1998).
11. J. Y. Chen, F. Saeki, B. J. Wiley, H. Cang, M. J. Cobb, Z. Y. Li, L. Au, H. Zhang, M. B. Kimmey, X. D. Li, and Y. N. Xia, "Gold nanocages: bioconjugation and their potential use as optical imaging contrast agents," Nano Lett. **5**, 473-477 (2005).

12. C. L. Nehl, H. W. Liao, and J. H. Hafner, "Optical properties of star-shaped gold nanoparticles," Nano Lett. **6**, 683-688 (2006).
13. Marek Piliarik and Jiří Homola, "Surface plasmon resonance (SRR) sensors: approaching their limits?" Opt. Express **17**, 16505-16517 (2009).
14. J. B. Pendry, A Holden, D. Robbins, and W. Stewart, "Magnetism from conductors and enhanced nonlinear phenomena," IEEE Trans. Microwave Theory Tech. **47**, 2075-2084 (1999).
15. V. M. Shalaev, W. S. Cai, U. K. Chettiar, H. K. Yuan, A. K. Sarychev, V. P. Drachev, and A. V. Kildishev, "Negative index of refraction in optical metamaterials," Opt. Lett. **30**, 3356-3358 (2005).
16. F. M. Wang, H. Liu, T. Li, S. N. Zhu, and X. Zhang, "Omnidirectional negative refraction with wide bandwidth introduced by magnetic coupling in a tri-rod structure," Phys. Rev. B **76**, 075110 (2007).
17. J. X. Cao, H. Liu, T. Li, S. M. Wang, T. Q. Li, S. N. Zhu, and X. Zhang, "Steering polarization of infrared light through hybridization effect in a tri-rod structure," J. Opt. Soc. Am. B **26**, B96-B101 (2009).
18. M. A. Ordal, L. L. Long, R. J. Bell, S. E. Bell, R. R. Bell, R. W. Alexander, Jr., and C. A. Ward, "Optical properties of the metals Al, Co, Cu, Au, Fe, Pb, Ni, Pd, Pt, Ag, Ti, and W in the infrared and far infrared," Appl. Opt. **22**, 1099 (1983).
19. H. Liu, D. A. Genov, D. M. Wu, Y. M. Liu, J. M. Steele, C. Sun, S. N. Zhu, and X. Zhang, "Magnetic plasmon propagation along a chain of connected subwavelength resonators at infrared frequencies," Phys. Rev. Lett. **97**, 243902 (2006).
20. Na Liu, Hui Liu, Shining Zhu, and Harald Giessen, Nature Photon. **3**, 157 (2009).